\documentclass{elsart5p}

\usepackage{graphics}
\usepackage{graphicx}
\usepackage{amssymb}
\begin{document}

\begin{frontmatter}

\title{Spin-size disorder model for granular superconductors with charging effects }

\author[AA]{Enzo Granato \corauthref{PPP}} and \ead{enzo@las.inpe.br}  \author[BB]{Giancarlo Jug}

\address[AA]{Laborat\'orio Associado e Sensores e Materiais,
Instituto Nacional de Pesquisas Espaciais, 12227-010 S\~ao Jos\'e
dos Campos, SP Brazil}
\address[BB]{Dipartimento di Fisica e Matematica, Universit\`a dell'Insubria, Via
Valleggio 11, 22100 Como, Italy \\
CNISM -- Unit\`a di Ricerca di Como and INFN -- Sezione di Pavia,
Italy }

\corauth[PPP]{Corresponding author. Tel/Fax: +55 12 39456717}

\begin{abstract}
A quantum pseudo-spin model with random spin sizes is introduced to
study the effects of charging-energy disorder on the superconducting
transition in granular superconducting materials. Charging-energy
effects result from the small electrical capacitance of the grains
when the Coulomb charging energy is comparable to the Josephson
coupling energy. In the pseudo-spin model, randomness in the spin
size is argued to arise from the inhomogeneous grain-size
distribution. For a particular bimodal spin-size distribution, the
model describes percolating granular superconductors. A mean-field
theory is developed to obtain the phase diagram as a function of
temperature, average charging energy and disorder.
\end{abstract}


\end{frontmatter}

\section{ Introduction}

Quenched disorder is obiquitous in condensed-matter systems,
sometimes determining entirely new physical properties and
phenomena. For the theoretical explanation of many experimental
findings, models of classical and quantum spin systems are employed
and - typically - disorder is introduced through a spatial variation
of the exchange couplings, of the direction of the axial
anisotropies, and of the sign of the couplings. Theoretical and
experimental studies of model and real physical systems with these
kinds of disorder have lead to much new physics. Much discussed have
been the glassy phases in magnetic \cite{sigfur96,wfsl95,sim}, and
superconducting materials \cite{fisher}.

In superconductors, specially with the advent of high-$T_c$ ceramic
superconductors, the role of disorder has  become central in the
discussion of the physical properties of real materials
\cite{sim,fisher}. Many materials are naturally microstructured and
granularity characterises the mesoscopic structures of most systems,
leading to a phase diagram often displaying a
superconductor-insulator transition at zero temperature due to the
charging energy of the grains. As first pointed out by Abeles
\cite{abeles}, when the grain charging energy arising from the
charge $Q$ and capacitance $C$ of the grain, $E_c = Q^2/ 2C \sim (2
e)^2/d$ ($d$ is the grain diameter and $e$ the electronic charge) is
larger than the Josephson-coupling energy $E_o$ between nearest
neighbors grains, phase coherence is destroyed due to zero-point
quantum fluctuations of the local phase of the superconducting order
parameter \cite{doniach}. Granularity can also be realized in a
controlled manner in artificially fabricated Josephson-junction
arrays of coupled superconducting grains \cite{fazio}, with a space
dimensionality $d$ less than $3$. Such granular systems can be
theoretically modelled by pseudo-spin systems, the 'spins'
representing with their states the few relevant quantum charge
states of the grains at relatively low temperatures. In such models
the pseudo-spin size is the same for all grains corresponding to the
assumption that the grain-size distribution is very narrow, or
alternatively, of negligible charging-energy disorder. A well known
model of this type is the pseudo-spin-one model introduced by de
Gennes and studied in different works \cite{sim,fazekas,gc93,kopec}
where only charge states $-1,0,1$ are allowed, corresponding to
$S=1$. However, realistic systems may contain different kinds of
disorder, such as a spatial distribution of Josephson couplings
between grains or/and a distribution of grain sizes, which leads to
disorder in the grain electrical capacitances and charging energies.

Studies of the effects of disorder in the electrical capacitance or
charging energy of the grains have appeared recently
\cite{mancini,alsaid1,alsaid2}. Within a mean-field approximation
\cite{mancini}, charging energy disorder widens the extent of the
superconducting phase at the expense of the insulating one. This
behavior is consistent with results for the superconductor-insulator
transition from quantum Monte Carlo simulations in two-dimensional
models of Josephson-junction arrays \cite{alsaid2} with disorder in
the diagonal capacitance matrix. Earlier calculations for the
related boson Hubbard model with disorder in the onsite Coulomb
repulsion \cite{cha} are also consistent with a decrease in the
phase coherence threshold and moreover suggest a different
universality class from the non-disordered case \cite{cha91}.

In this work, we consider the effects of charging energy disorder in
granular superconducting materials within the pseudo-spin approach.
By considering a spatial distribution of the grain sizes, which
leads to local charging energy disorder, a quantum pseudo-spin model
with random on-site spin sizes can be constructed.
A mean-field theory is developed to obtain
the phase diagram as a function of temperature, average charging
energy and disorder.

Spin-size disorder models have seldom been considered in the
literature. In the context of classical spin models, spin-size
disorder can be readily turned into exchange-coupling disorder; the
physics of quantum systems with spin-size disorder, however, appears
to have not been investigated in such depths. It has been considered
mostly within the one-dimensional systems: dilution of a quantum
spin-$\frac{1}{2}$ 2-ladder was studied by Sigrist and Furusaki
\cite{sigfur96} whilst the general problem of a quantum spin chain
with random $S$ as well as random $J$ was considered by Westerberg
{\it et. al} \cite{wfsl95} within  a real-space
renormalization-group method showing that these systems belong to a
different universality class of disordered spin systems.

\section{Pseudo-spin model with random spin sizes}

The Hamiltonian for a set of superconducting grains coupled by the
Josephson energy can be written as the  sum of the Coulomb charging
energy and the Josephson-coupling energy \cite{sim,doniach,fazekas}
\begin{equation}
 H_{gs} =\frac{1}{2}\sum_{i,j}
C^{-1}_{ij} Q_i Q_j -\sum_{<ij>} E_{ij}\cos(\theta_i-\theta_j),
\end{equation}
where $Q_i$ is the net charge on the superconducting grain at site
$i$ and $\theta_i$ is the phase of the local superconducting order
parameter $\psi_i$. $C_{ij}$ is the electrical capacitance matrix
and $E_{ij}$ is the Josephson coupling between nearest-neighbor
grains. The charge $Q_i$ in each grain can be expressed in terms  of
the excess number of Cooper pairs $n_i$ as $Q_i=2 e n_i$.
Considering a diagonal capacitance matrix $C_{ij} = C_i \delta_{ij}$
and uniform Josephson-coupling energies $E_{ij} =E_o$, leads to the
self-charging model
\begin{equation}
 H_{gs}= 2\sum_i U_i n_i^2
-E_o\sum_{<ij>} \cos(\theta_i-\theta_j), \label{jja}
\end{equation}
where $U_i=e^2/C_i$ are the charging energies of the grains.  The
number operator $n_i$ is conjugate to the phase $\theta_i$,
satisfying the commutation relation
\begin{equation}
[n_i,\theta_j]=-i\delta_{ij}, \label{jjacomm}
\end{equation}
and can be written as $n_i=-i\partial/\partial\theta_i$, having
integer eigenvalues $ 0, \pm 1, \pm 2 ...$. The model of Eq.
(\ref{jja}) can also be regarded as a boson Hubbard model
\cite{cha,cha91} where the charging energy represents the onsite
Coulomb repulsion of bosons and the Josephson coupling  represents
the hopping term.

When the charging energies are uniform $U_i=U_o$, a  pseudo-spin one
model for the Hamiltonian of Eq. (\ref{jja}) can be constructed by
truncating the basis vectors of the number operators $| n_i > $ to
$|0>$ and $| \pm 1>$, corresponding to the lowest charging energy
states. Identifying the charging states, $| n_i>$, as the
eigenstates of  $S^z$ for spin $S=1$, writting the second term of
Eq. (\ref{jja}) in terms of $e^{\pm i\theta_i}$  and making the
correspondence
\begin{equation}
e^{i\theta_i}\to S_i^+/\sqrt{2}, \qquad e^{-i\theta_j}\to
S_j^-/\sqrt{2} ,\qquad  n_i \to  S_i^z, \label{pseudomap}
\end{equation}
one obtains the de Gennes mapping to a $S=1$ pseudo-spin model with
single-ion spin-anisotropy \cite{sim,fazekas}
\begin{equation}
H_{S=1} =2U_o\sum_i (S_i^z)^2 - \frac{E_o}{4} \sum_{\langle ij
\rangle }(S_i^+S_j^-+S_j^+S_i^-), \label{gennes}
\end{equation}
with $S_i^+=S_i^x+iS_i^y$. The factor $1/\sqrt{2}$ in front of
$S^{\pm}$ in Eq. (\ref{pseudomap}) takes into account the length of
the pseudo spin $\sqrt{S(S+1)}$. In terms of this spin model, the
superconductor to insulator transition in the original granular
superconductor corresponds to a zero temperature transition where
the ferromagnetic order of the spins in the $x y$ plane is destroyed
by quantum fluctuations when the $S^z$ component is confined to zero
for an increasing ratio of $U_o/E_o$. Although this mapping is not
exact \cite{das}, the $T=0$ critical behavior observed in the
original phase model of Eq. (\ref{jja}) without disorder and the
spin model is the same \cite{cha,glaus} as found in numerical
calculations in one dimension. For higher dimensions, results from
the same mean-field approximation applied to both models also agree
\cite{fazekas}.

To consider the main effects of disorder in the charging energies
$U_i$, we generalize the above approximate mapping to a pseudo-spin
model with randomness in the spin-size values $S_i$. Since for a
given charging energy (or temperature fluctuation), low values of
$U_i$ correspond to charging states with higher $n_i$, it seems
reasonable to use a truncation scheme which identifies the charging
states $|n_i>$ and $|n_j>$ with the $S^z$-eigenstates of the spins
of different sizes $S_i$ and $S_j$, such that the corresponding
maximum charging energies $2 U_i {\cal S}_i^2$ and $ 2U_j {\cal
S}_j^2$ of the grains are both comparable to the same truncation
energy (here ${\cal S}_i=\sqrt{S_i(S_i+1)}$ is the length of the
spin ${\bf S}_i$). This (approximate) mapping leads to the effective
spin Hamiltonian
\begin{equation}
H = D \sum_i \frac{1}{{\cal S}_i^2}(S_i^z)^2 + J \sum_{\langle ij
\rangle }\frac{1}{{\cal S}_i {\cal S}_j}(S_i^+S_j^-+S_j^+S_i^-),
\label{Haniso}
\end{equation}
with randomness in the spin-size values $S_i$. Here,  $J = -E_o/2$,
$D = 2 \bar U \bar {\cal S}^2$  where $\bar U$ is the average value
of $U_i$ and $\bar {\cal S}$ the average value of ${\cal S}_i$. The
spin sizes $S_i$ are restricted to take only integer values
$0,1,2,3, ...$ in this mapping. In Eq. (\ref{Haniso}), randomness in
the spin-size also leads to randomness in the local single-ion
anisotropy parameter, but these random variables are only correlated
at the same site.

A typical spin-size distribution we shall be considering is
\begin{equation}
P(S_i)=x\sum_{L=\pm 1}\delta(S_i-S_o-L)+(1-2x)\delta(S_i-S_o),
\label{distr}
\end{equation}
where three spin-size values $S_o, S_o +1$ and $S_o -1$ are mixed
with average value $S_o$ and a concentration $x$, which is a measure
of the disorder. It is also convenient to rewrite the spin model of
Eq. (\ref{Haniso}) in terms of normalized spins $\tilde S_i =
S_i/{\cal S}_i$
\begin{equation}
H = D \sum_i ({\tilde S_i}^z)^2 + J \sum_{\langle ij \rangle
}({\tilde S_i}^+{ \tilde S_j}^-+{\tilde S_j}^+{\tilde S_i}^-).
\label{Hanison}
\end{equation}
In this latter form, a bimodal distribution of spin values can be
considered:
\begin{equation}
P(\tilde S_i)=x\delta(\tilde S_i-\tilde S_1)+(1-x)\delta(\tilde
S_i-\tilde S_2), \label{distrd}
\end{equation}
which in the special case $\tilde S_1=0$ corresponds to the dilution
of a spin-$S_2$ system or to percolating granular superconductors
\cite{john} in the presence of charging effects.

\section{ Mean-Field Theory }

For high enough space dimensions $d$, reasonable results can be
obtained from simple mean-field approximations (MFA). The
MFA replaces all variables around a given small cluster
with their average value, or order parameter, and then use the
resulting approximate Hamiltonian to evaluate the order parameter
itself. Since the spin sizes are random, one must carry out the
averaging procedure with respect to $P(S_i)$ as well and we denote
this by $[...]_{ave}$.

The simplest cluster is a single spin on a site. With ${\bf
M}=[<{\bf S}_i>]_{ave}$ the mean field, this yields a
self-consistent equation for the order parameter ${\bf M}$:

\begin{equation}
{\bf M}=[\langle {\bf S}_i \rangle]_{ave}=\left[ \frac{1}{Z_{MFA}}
Tr_{{\bf S}_i} {\bf S}_i e^{-\beta{\bar H}_{MFA}} \right]_{ave},
\label{mfa1}
\end{equation}
where $\bar H_{MFA}$ is the mean-field Hamiltonian and $\beta =
1/k_B T$. For the anisotropic Hamiltonian of  Eq. (\ref{Haniso}), we
assume that the ferromagnetic ordering takes place in the $xy$ plane
of the anisotropy, say $[<S_i^x>]_{ave}=M$. Then, in the MFA we replace
\cite{sim}

\begin{equation}
S_i^xS_j^x+S_i^yS_j^y\to \langle S_i^x\rangle S_j^x +S_i^x\langle
S_j^x\rangle-\langle S_i^x\rangle\langle S_j^x\rangle, \label{amfa1}
\end{equation}
by virtue of the fact that $<S_i^y>=0$. Near the critical curve, the
MFA Hamiltonian then becomes:

\begin{equation}
{\bar H}_{MFA}= D\sum_i(({\tilde S_i}^z)^2-\lambda {\tilde
S_i}^x)+O({\tilde M}^2), \label{amfa2}
\end{equation}
where $\lambda= 2z|J|{\tilde M}/D$ is a dimensionless expansion parameter.

Since we are interested in the phase transition line only, we can
assume $\tilde M \equiv M/\mathcal{S}$ small and use a first-order
perturbation expansion to solve Eq. (\ref{mfa1}) for arbitrary spin
$S$. In a now single-site problem, we reabsorb a factor ${\cal
S}^{-2}$ into $D$ and $J$. The unperturbed states $|m>$ are
eigenstates of $S^z$ and thus also of the unperturbed Hamiltonian
$H_o=D(S^z)^2$, with $m=-S, -S+1, ..., S-1, S$. We have
\begin{equation}
\langle S^x \rangle=\frac{ \sum_m e^{-\beta E_m}\langle \psi_m\vert
S^x \vert\psi_m\rangle }{ \sum_m e^{-\beta E_m} }, \label{def}
\end{equation}
where to first-order
$E_m=Dm^2+\langle m\vert -\lambda DS^x \vert m\rangle=Dm^2$ and

\begin{eqnarray}
\vert\psi_m\rangle&=&\vert m\rangle+{\sum_{m',m}}^{\prime} \ \frac{
\langle m'\vert -\lambda DS^x\vert m\rangle }{ E_m^0-E_{m'}^0 }\vert
m'\rangle
\nonumber \\
&=&\vert m\rangle-\lambda {\sum_{m',m}}^{\prime} \ \frac{ \langle
m'\vert S^x\vert m\rangle }{ m^2-m'^2 }\vert m'\rangle. \label{1st}
\end{eqnarray}
Here, the prime on $\sum^{\prime}$ means $m'\not=  m$. These can
then be inserted in Eq. (\ref{def}) to get, keeping only first order
terms in $\lambda$:

\begin{equation}
\langle S^x \rangle=\frac{2}{Z_0}{\sum_{m',m}}^{\prime}\ e^{-\beta
Dm^2} \frac{\langle m\vert S^x \vert m'\rangle\langle m'\vert S^x
\vert m\rangle} {m'^2-m^2}\lambda, \label{amfa3}
\end{equation}
where $Z_0 = \sum_{m=-S}^S e^{-\beta Dm^2}$. Using

\begin{eqnarray}
\langle m\vert S^x \vert m'\rangle
&=&\frac{1}{2}(A_{m-1}^S\delta_{m',m-1}
+A_m^S\delta_{m',m+1}) \nonumber \\
A_m^S &=& \sqrt{(S-m)(S+m+1)}, \label{matrix}
\end{eqnarray}
we get

\begin{equation}
 \langle S^x \rangle =  \frac{\lambda}{2Z_0}\sum_m \{
\frac{(A_m^S)^2}{(m+1)^2-m^2} \\
 + \frac{(A_{m-1}^S)^2}{(m-1)^2-m^2}\} e^{-\beta Dm^2}
\end{equation}
which, using Eq. (\ref{mfa1}), gives implicitly the critical
temperature in the MFA

\begin{eqnarray}
\nonumber  \frac{D}{2z\vert J\vert} = & &  \big[
 \frac{1}{\sum_{m} e^{-Dm^2/(S(S+1) k_BT_c})} \\
& & \sum_{m=-S}^S\frac{S(S+1)+m^2}{1-4m^2}e^{-Dm^2/(S(S+1) k_BT_c)}
\big]_{ave} . \label{amfa7}
\end{eqnarray}

One can check that for  $S=1$ and in absence of disorder the MFA
transition line from Eq. (\ref{amfa7}) agrees with the result
obtained for the de Gennes model of Eq. (\ref{gennes}) in the same
approximation \cite{sim}. Fig. 1 shows the phase boundaries between
the superconducting and normal phases obtained in absence of
disorder for different values of uniform spin sizes $S=1$, $S=2$ and
$S=5$. The critical value $E_o/U_o$ for the superconductor-insulator
transition at $T=0$ remains unchanged and only at much higher
temperatures there is a deviation in the transition lines when
considering larger spin approximations. This is in agreement with MF
calculations using truncated basis vectors in the phase model of Eq.
(\ref{jja}) for increasing number of states \cite{fazekas}.

Our present calculations generalize the MFA to arbitrary spin sizes
and allow to obtain the transition line in the case of a random spin
size distribution  by simply averaging with respect to $P(S)$ in Eq.
(\ref{amfa7}). The consequences of spin-size disorder depend very
much now on the chosen distribution, but it is expected that in
general within a simple MF approximation, the phase transition of
the homogeneous system is preserved with at most a modification of
the phase boundary line. Fig. 2a shows the transition lines for the
spin-size distribution of Eq. (\ref{distr}) with $S_o=2$ and
different values of the disorder parameter $x$. For increasing
disorder the critical value for phase coherence $E_o/\bar U$
decreases, increasing the extent of the superconducting phase at low
enough temperatures. This behavior is in qualitative agreement with
MF calculations \cite{mancini} in the phase model of Eq.
(\ref{jja}).

\begin{figure}
\includegraphics[ bb= 0cm 4cm  18cm 11.5cm,width=7 cm]{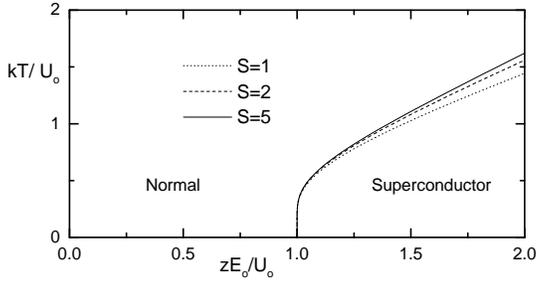}
\caption{Phase boundary without disorder obtained from the
pseudo-spin model with different uniform spin values $S_i=1, 2$ and
$S_i=5$. }
\end{figure}

On the other hand, a different behavior is expected when there is
dilution of superconducting grains. Fig. 2b shows the transition
lines for the spin-size distribution of Eq. (\ref{distrd}) with
$\tilde S_1=0$, $S_2=2$, corresponding to a dilution of grains for
different concentrations $x$. In  this case, disorder decreases the
extent of the superconducting phase. For large values of $x$,
corresponding to the system below the percolation threshold, the
superconducing phase should disappear. However, an improved  MF
approximation is required to describe this behavior. One possible
approach is the MF renormalizaton-group method \cite{indekeu} used
previously to study the de Gennes model of Eq. (\ref{gennes}) in
absence of disorder \cite{gc93}.

\begin{figure}
\includegraphics[ bb= 0cm 4cm  18cm  21cm,width=7 cm]{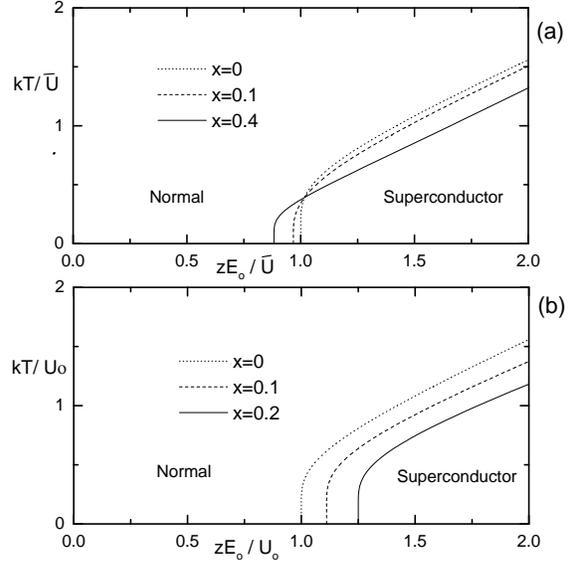}
\caption{Phase boundary with disorder obtained from the pseudo-spin
model with random spin sizes:  a)  with charging energy disorder,
corresponding to disordered spins values giving by the probability
distribution of Eq. (\ref{distr}), with average spin value $S_o=2$
and different $x$ ; b) with dilution of grains, corresponding to the
disordered spin values with the probability distribution of Eq.
(\ref{distrd}), with $\tilde S_1=0$ and $S_2=2$ and different $x$}
\end{figure}

\section{Conclusions}

We have introduced a  quantum pseudo-spin model with random spin
sizes to model the effects of charging-energy disorder in granular
superconducting materials. Randomness in the spin size is argued to
arise from the inhomogeneous grain-size distribution. For a
particular bimodal spin-size distribution, the model describes
percolating granular superconductors. A mean-field theory has been
developed to obtain the phase diagram as a function of temperature,
average charging energy and disorder. The results are qualitatively
consistent with previous mean-field calculations in the phase-number
representation. The pseudo-spin model should provide a useful
framework to study the critical behavior and universality classes in
presence of strong charging-energy disorder.

\section{Acknowledgments}

E.G. was suported by   by Funda\c c\~ao de Amparo \`a Pesquisa do
Estado de S\~ao Paulo (FAPESP) under Grant No. 07/08492-9. G.J.
acknowledges FAPESP (Grant No. 07/ 04006-2) for supporting a visit
to the Instituto Nacional de Pesquisas Espaciais, where part of this
work was done.


\begin{thebibliography}{99}

\bibitem{sigfur96} M. Sigrist and A. Furusaki, J. Phys. Soc. Jpn {\bf
65}, 2385 (1996).

\bibitem{wfsl95} E. Westerberg, A. Furusaki, M. Sigrist and P.A. Lee,
Phys. Rev. Lett. {\bf 75}, 4302 (1995); {\it ibid.} Phys. Rev. B
{55}, 12578 (1997).

\bibitem{sim} E. Sim\'anek, {\em Inhomogeneous Superconductors: Granular
and Quantum Effects} (Oxford UP, New York 1994)

\bibitem{fisher} D.S. Fisher, M.P.A. Fisher, and D. Huse, Phys. Rev.
B {\bf 43}, 130 (1991).

\bibitem{abeles} B. Abeles, Phys. Rev. B {\bf 15}, 2828 (1977).

\bibitem{doniach} S. Doniach, Phys. Rev. B {\bf 24}, 5063 (1981).

\bibitem{fazio} R. Fazio and H. van der Zant, Phys. Rep. {\bf 555},
235 (2001).

\bibitem{fazekas} P. Fazekas, B. Muhlschlegel, and M. Schroter, Z.
Phys. B {\bf 57}, 193 (1984).

\bibitem{gc93} E. Granato and M.A. Continentino, Phys. Rev. B {\bf 48}, 15977
(1993).

\bibitem{kopec} T.K. Kope\'c, Phys. Rev. B {\bf 69}, 054504 (2004).

\bibitem{mancini} F.P. Mancini, P. Sodano and A. Trombettoni, Phys. Rev. B
{\bf 67}, 014518 (2003).

\bibitem{alsaid1} W.A. Al-Saidi and D. Stroud, Phys. Rev. B {\bf 67},
024511 (2003).

\bibitem{alsaid2} W.A. Al-Sadi and D. Stroud, Physica C {\bf 402},
216 (2004).

\bibitem{cha} M.-C. Cha and S.M. Girvin, Phys. Rev. B {\bf 49},
9794 (1994).

\bibitem{cha91} M.-C. Cha, M.P.A. Fisher, S.M. Girvin, M. Wallin,
and A.P. Young, Phys. Rev. B {\bf 44}, 6883 (1991).

\bibitem{john} S. John and T.C. Lubensky, Phys. Rev. Lett. {\bf 55}, 1014
(1985).

\bibitem{indekeu} J.O. Indekeu, A. Maritan, and A. Stella, J. Phys.
A {\bf 15}, L291 (1982).

\bibitem{das} D. Das and S. Doniach, Phys. Rev. B {\bf 60}, 1261
(1999).

\bibitem{glaus} U. Glaus, Physica {\bf 141}A, 295 (1987).

\end{thebibliography}
\end{document}